\documentclass[nofootinbib,twocolumn,preprintnumbers,longbibliography]{revtex4-1}

\usepackage{amsmath}
\usepackage{amsfonts}
\usepackage{microtype}
\usepackage{hyperref}
\usepackage[noabbrev]{cleveref}
\usepackage{color}
\usepackage{graphicx}

\newcommand{\dd}{\mathrm{d}}
\newcommand{\sdg}{\sqrt{-g}}
\newcommand{\mn}{{\mu\nu}}
\newcommand{\ab}{{\alpha\beta}}

\newcommand{\Mp}{M_\mathrm{Pl}}

\newcommand{\nbl}{n_{B-L}}

\makeatletter
\DeclareRobustCommand{\rcite}[1]{%
  \rcite@aux#1,\@nil{#1}%
}
\def\rcite@aux#1,#2\@nil#3{%
  \if\relax#2\relax
    Ref.~\cite{#3}%
  \else
    Refs.~\cite{#3}%
  \fi
}
\makeatother

\begin{document}

\title{Baryogenesis in Lorentz-violating gravity theories}
\author{Jeremy Sakstein}
\email{sakstein@physics.upenn.edu}
\author{Adam R. Solomon}
\email{adamsol@physics.upenn.edu}
\affiliation{Center for Particle Cosmology, Department of Physics and Astronomy, University of Pennsylvania, 209 S. 33rd St., Philadelphia, PA 19104, USA}

\begin{abstract} 
Lorentz-violating theories of gravity typically contain constrained vector fields. We show that the lowest-order coupling of such vectors to $\mathrm{U}(1)$-symmetric scalars can naturally give rise to baryogenesis in a manner akin to the Affleck-Dine mechanism. We calculate the cosmology of this new mechanism, demonstrating that a net $B-L$ can be generated in the early Universe, and that the resulting baryon-to-photon ratio matches that which is presently observed. We discuss constraints on the model using solar system and astrophysical tests of Lorentz violation in the gravity sector. Generic Lorentz-violating theories can give rise to the observed matter-antimatter asymmetry without violating any current bounds.
\end{abstract}


\maketitle

\section{Introduction}

Why is there so much more matter than antimatter? One most likely cannot appeal to initial conditions, as these would be washed away by inflation. The standard model can provide such an asymmetry during the electroweak phase transition, but cannot produce enough to accommodate observations \cite{Morrissey:2012db}. It seems probable, then, that a dynamical generation mechanism, or \emph{baryogenesis}, arises from new physics beyond the standard model.\footnote{For reviews of baryogenesis, see, e.g., \rcite{Trodden:1998ym,Trodden:2004mj,Riotto:1999yt,Dine:2003ax,Cline:2006ts,Morrissey:2012db,Allahverdi:2012ju}.}

In this paper we point out that Lorentz violation might play a key role in this new physics. While Lorentz invariance is extraordinarily well tested in the matter sector, the possibility of gravitational Lorentz violation remains relatively unconstrained. If boosts are broken but rotational invariance is maintained---i.e., if gravity picks out a preferred rest frame---then the low-energy physics is described by Einstein-\ae ther theory, a vector-tensor theory in which the vector field is constrained to have a fixed, timelike norm. This is the general effective field theory when boosts are broken \cite{ArmendarizPicon:2010mz}; for example, a special case of Einstein-\ae ther arises in the low-energy limit of Ho\v{r}ava-Lifschitz gravity \cite{Horava:2009uw,Blas:2010hb}, a putative UV completion of general relativity which relies on the existence of a preferred foliation.

We demonstrate that if a $\mathrm{U}(1)_{B-L}$ scalar couples to the vector of Einstein-\ae ther theory, then the lowest-order interactions between the two can lead to baryogenesis. This operates in a manner qualitatively similar to Affleck-Dine baryogenesis \cite{Affleck:1984fy} or the recent model of \rcite{Sakstein:2017lfm}, in which the $\mathrm{U}(1)_{B-L}$ symmetry is broken at early times due to a tachyonic mass proportional the Hubble parameter $H$ appearing in the effective scalar potential.\footnote{In the former model, this arises due to a coupling of the scalar to the inflaton, while in the latter, it arises from a Weyl coupling to dark matter.} In purely metric theories this is difficult to achieve, as $H$ is not a spacetime scalar. Breaking boosts cures this difficulty, and indeed in Einstein-\ae ther $H$ is simply proportional to the divergence of the timelike vector field. As Einstein-\ae ther is the most general low-energy effective theory for broken boosts, our conclusion can be stated as follows: if the Universe contains a $\mathrm{U}(1)_{B-L}$ scalar with softly broken symmetry and spontaneous Lorentz violation, a working baryogenesis mechanism comes for free.

This paper is organized as follows. In \cref{sec:model} we introduce the model of scalar-\ae ther baryogenesis and discuss known constraints on the theory. In \cref{sec:cosmo} we derive the cosmology and verify that this model can yield the observed baryon-to-photon ratio with sensible parameters, and we conclude in \cref{sec:conc}.

\section{Model}
\label{sec:model}

The model we will consider is the constrained vector (or ``\ae ther") $u^\mu$ of Einstein-\ae ther theory coupled to a new $\mathrm{U}(1)_{B-L}$ scalar $\phi$. At leading order, the most general action we can write down is
\begin{align}\label{eq:action}
S&=\int\dd^4x\sdg\bigg[\frac{\Mp^2}{2} R-\mathcal{K}^\mn_{\alpha\beta}\nabla_\mu u^\alpha\nabla_\nu u^\beta+\kappa\left(u^\mu u_\mu+m^2\right)\nonumber\\
&\hphantom{{}=}-\partial_\mu\phi\partial^\mu\phi^\dagger-m_\phi^2|\phi|^2-\frac{\lambda}{2}|\phi|^4-\frac{\varepsilon}{4}\phi^4-\frac{\varepsilon^\dagger}{4}{\phi^\dagger}^4\nonumber\\
&\hphantom{{}=}+\frac\alpha3|\phi|^2\nabla_\mu u^\mu\bigg],
\end{align}
where
\begin{equation}
\mathcal{K}^\mn_{\alpha\beta}= c_1g^{\mu\nu}g_{\ab}+c_2\delta^\mu_\alpha\delta^\nu_\beta+c_3\delta^\mu_\beta\delta^\nu_\alpha+c_4u^\mu u^\nu g_\ab
\end{equation}
is the most general kinetic term for the Lorentz-violating vector $u^\mu$, and $\kappa$ is a Lagrange multiplier that ensures that the vector is timelike and of fixed norm, $u^\mu u_\mu=-m^2$. We have included some $\mathrm{U}(1)_{B-L}$-violating terms proportional to $\varepsilon$ in order to generate a net $B-L$. The last line is the leading-order interaction one can write down between $\phi$ and $u^\mu$ given the symmetries.\footnote{Couplings between a scalar and $\nabla_\mu u^\mu$ were first introduced in \rcite{Donnelly:2010cr}, and have also been considered in, e.g., \rcite{Blas:2011en,Barrow:2012qy,Solomon:2013iza,Sandin:2012gq,Ivanov:2014yla}. One could also consider a term $|\phi|^2u^\mu u_\mu$, but this can always be absorbed into the mass and Lagrange multiplier when the vector is on-shell.} This is the general low-energy effective theory with $\phi$, broken boosts, and softly-broken $\mathrm{U}(1)_{B-L}$.

The coupling between $u^\mu$ and $\phi$ can give rise to baryogenesis using a mechanism akin to (but distinct from) that of Affleck and Dine \cite{Affleck:1984fy} or similar generalizations \cite{Sakstein:2017lfm}. In particular, the effective potential for $\phi$ is
\begin{align}\label{eq:Veff1}
V_\mathrm{eff}(\phi,\phi^\dagger)&=\left(m_\phi^2-\frac13\alpha\nabla_\mu u^\mu\right)|\phi|^2\nonumber\\
&\hphantom{{}=}+\frac{\lambda}{2}|\phi|^4+\frac{\varepsilon}{4}\phi^4+\frac{\varepsilon^\dagger}{4}{\phi^\dagger}^4
\end{align}
so that $\nabla_\mu u^\mu$ acts as a tachyonic mass term for $\phi$. We can see this explicitly by considering a homogeneous and isotropic cosmological setting, so that the Universe is described by a Freidmann-Lema\^{i}tre-Robertson-Walker metric (in cosmic time)
\begin{equation}
\dd s^2=-\dd t^2+ a^2(t)\dd \vec{x}^2.
\end{equation}
The on-shell condition $u_\mu u^\mu=-m^2$ and symmetry imply that the vector must be of the form \cite{Carroll:2004ai,Lim:2004js,Carruthers:2010ii}
\begin{equation}
u^\mu=\left(m,\,0,\,0,\,0\right).
\end{equation}
In this case, the divergence of $u^\mu$ is $\nabla_\mu u^\mu=m\partial_t\ln\sqrt{-g}=3mH$, so that the effective potential becomes
\begin{equation}\label{eq:Veff-cosmo}
V_{\rm eff}(\phi,\phi^\dagger)=\left(m_\phi^2-\alpha mH\right)|\phi|^2+\frac{\lambda}{2}|\phi|^4+\frac{\varepsilon}{4}\phi^4+\frac{\varepsilon^\dagger}{4}{\phi^\dagger}^4.
\end{equation}

This potential leads to baryogenesis similarly to the well-known Affleck-Dine mechanism. In the early Universe, the $\mathrm{U}(1)_{B-L}$ symmetry is broken as the tachyonic mass term is more important than the bare mass, while at late times the symmetry is restored. During the broken-symmetry phase, the motion of the angular component of the scalar generates a net $B-L$ due to the symmetry-breaking terms $\varepsilon\phi^4+\varepsilon^\dagger{\phi^\dagger}^4$. When the symmetry is restored, the $B-L$ is stored in the field, and can be transferred to the standard model through sphaleron processes \cite{Harvey:1990qw}, although one must first transfer the asymmetry to left-handed standard model particles. The details of the transfer were discussed for models such as ours in \rcite{Sakstein:2017lfm}, where the neutrino portal \cite{Falkowski:2009yz,Macias:2015cna} was identified as one promising mechanism. We note that our model differs quantitatively from Affleck-Dine: the tachyonic mass scales like $mH$ rather than $H^2$, which can lead to novel and interesting new features. In what follows, we will calculate the cosmology of this model, paying special attention to the generation of a net $B-L$.

\subsection{Constraints}
\label{sec:constraints}

In this subsection, we briefly summarize observational and theoretical constraints on the parameters in our model. Most of these will apply to Einstein-\ae ther theory or to its coupling to a real scalar. We will use the notation $c_{12}=c_1+c_2$, $c_{123}=c_1+c_2+c_3$, etc.

Experimental constraints on Einstein-\ae ther theory tend to place upper bounds on the \ae ther vacuum expectation value (VEV) $m$, with the result $c_im^2\ll\Mp^2$ for generic values of the $c_i$ parameters. We note that any of these constraints can be weakened or removed entirely by tuning the $c_i$ parameters, although these tunings cannot all be done simultaneously. We refer the reader to Sec. V.D of \rcite{Solomon:2013iza} for a more comprehensive summary of constraints on the \ae ther.

The strongest constraints come from gravitational \u{C}erenkov radiation: high-energy cosmic rays could lose energy to subluminal \ae ther-graviton modes, leading to a degradation in cosmic ray propagation which has not been observed, constraining $m/\Mp<3\times 10^{-8}$ \cite{Elliott:2005va}. These constraints can be avoided by tuning the $c_i$ or allowing for superluminal propagation in \ae ther-graviton modes; since this is an explicitly Lorentz-breaking theory, superluminality may not be as deadly as one normally expects. The preferred-frame parameters $\alpha_{1,2}$ in the parametrized post-Newtonian formalism are modified by the \ae ther, constraining $m/\Mp<6\times10^{-4}$ in the absence of tuning $c_i$ \cite{Foster:2005dk,Jacobson:2008aj}. Note that the tuning which eliminates gravitational \u{C}erenkov radiation ($c_3=-c_1$, $c_2=c_1/(1-2c_1)$) also sets $\alpha_1=0$, so that the dominant constraint comes from $\alpha_2$, in which case the strongest constraint on $m$ is rather mild, $m/\Mp\lesssim10^{-2}$.

Under the assumption that $c_im^2\ll\Mp^2$, we are justified in ignoring the mixing with gravity \cite{Lim:2004js}, in which case there are a few constraints on the $c_i$ from flat space perturbation theory. In the vector sector, the absence of ghosts requires $c_1>0$ \cite{Lim:2004js}; coupling a scalar to $\nabla_\mu u^\mu$, as in this paper, does not modify the vector perturbations around flat space \cite{Solomon:2013iza}. The no-ghost condition for the spin-0 piece of $u^\mu$ is the same, while gradient stability requires $c_{123}>0$. Some authors require the spin-0 \ae ther mode to propagate subluminally, which would imply $c_{123}<c_1$ \cite{Lim:2004js}, although the scalar coupling relaxes this bound to $c_{123}<c_1+x$ for some $x>0$ \cite{Solomon:2013iza}. If we require the sound speed of tensors to be subluminal then we would require $c_{13}>0$ \cite{Lim:2004js}.

\begin{figure*}[ht]
{\includegraphics[width=0.45\textwidth]{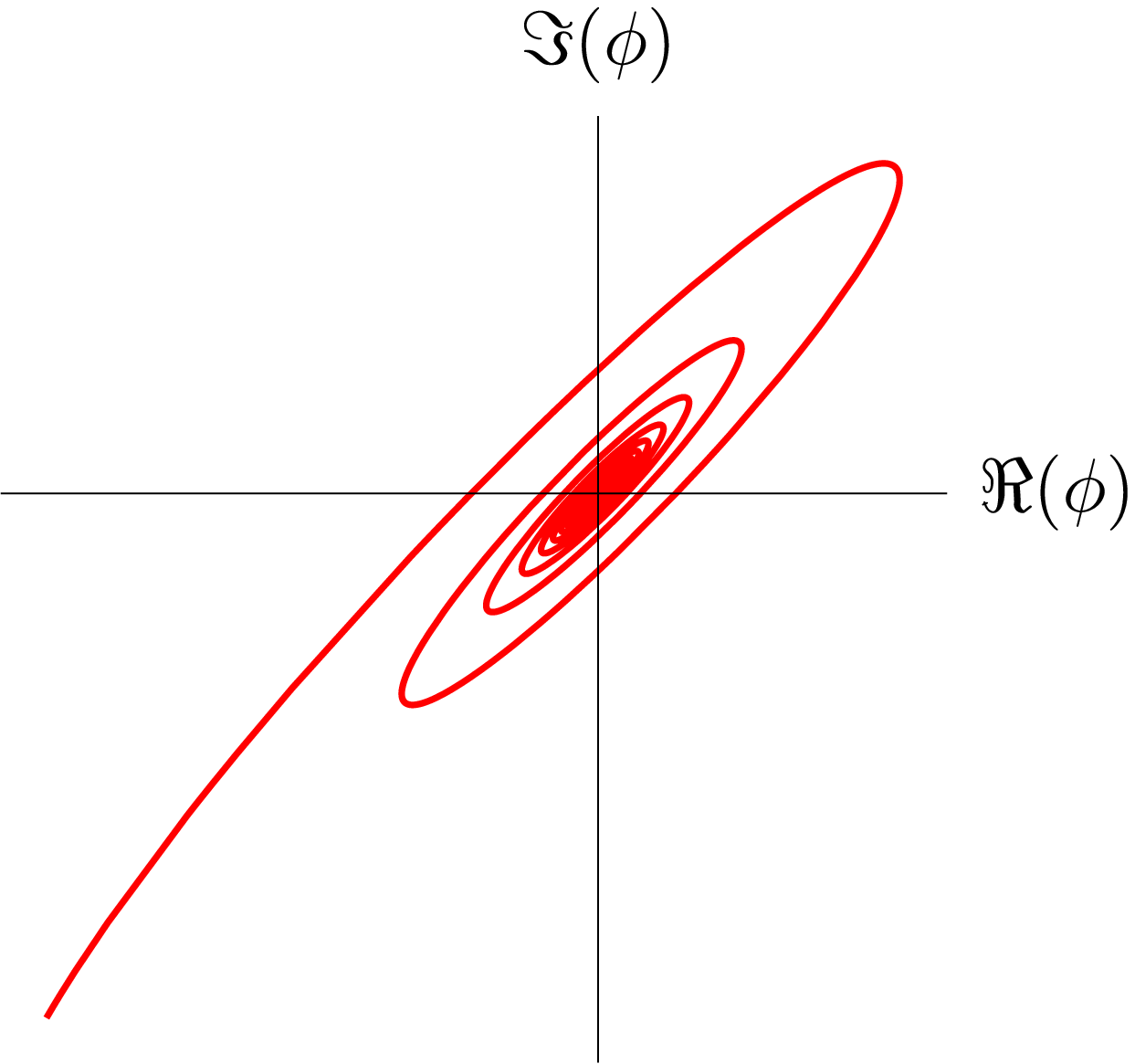}}
{\includegraphics[width=0.45\textwidth]{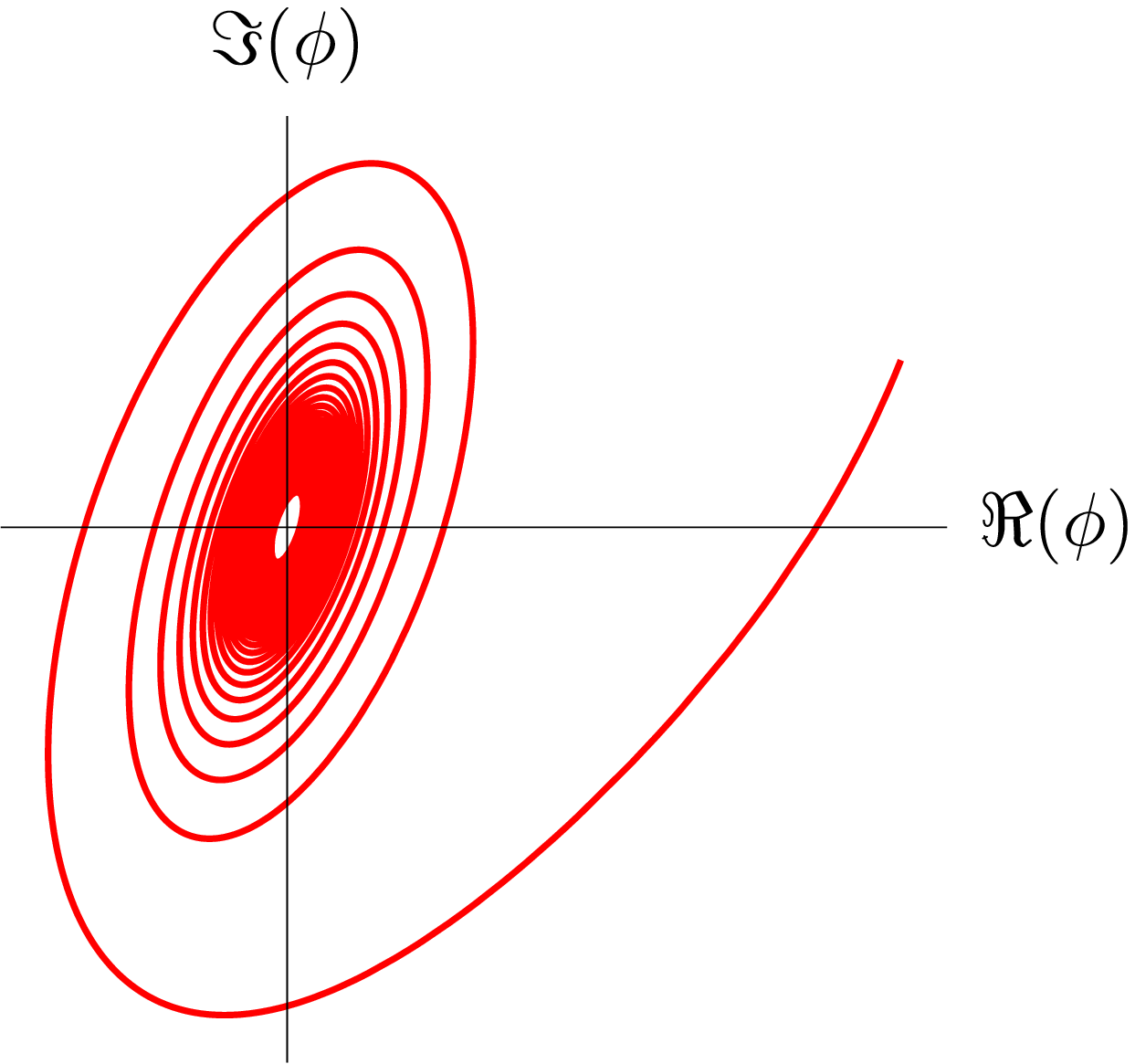}}
\caption{The motion of the complex scalar in field space. The left panel corresponds to a model with $m_\phi/\alpha m=1/2$, $\alpha m/\Mp=4\times10^{-7}$, and $T_R=10^{11}$ GeV.  The right panel has $m_\phi/\alpha m=7\times10^{-3}$, $\alpha m/\Mp=2\times10^{-6}$, and $T_R=10^{10}$ GeV.}
\label{fig:field}
\end{figure*}

The \ae ther-scalar coupling can lead to a gradient instability, placing an upper bound on $\alpha$ \cite{Solomon:2013iza}. Writing $\phi(x) = \frac{1}{\sqrt{2}} \rho(x) e^{i\theta(x)}$, the real scalar $\rho$ interacts with the \ae ther through a potential $V(\theta,\rho)=\frac12m_\phi^2\rho^2 + \frac14\lambda \rho^4 - \frac16\alpha\Theta\rho^2$, where $\Theta\equiv\nabla_\mu u^\mu$. Gradient stability around flat space requires\footnote{While the analysis of \rcite{Solomon:2013iza}, in contrast to our model, assumed a single real scalar, the field $\theta$ decouples from $\rho$ and $u^\mu$ around the background $\partial_\mu\bar\theta=0$. In principle a non-zero $\partial_\mu\bar\theta=c_\mu$ could modify the constraint by shifting the effective mass for $\rho$ fluctuations, $m_\phi^2\to m_\phi^2+c^2$.}
\begin{equation}
V_{\Theta\rho}^2 \leq 2c_{123}\left(V_{\rho\rho} + k^2\right),
\end{equation}
where $V_{\Theta\rho}=\partial_\Theta\partial_\rho V$ and $V_{\rho\rho} = \partial_\rho^2 V$ are evaluated at the background values of $\Theta$ and $\rho$. Applying this to our potential, we find that the constraint is trivially satisfied in the unbroken symmetry phase ($\rho=0$), while in the broken symmetry phase ($\rho=\bar\rho$) we have, in the $k\to0$ limit,\footnote{Strictly speaking the $k\to0$ limit is not physical, as we will be dealing with cosmological spacetimes, which only resemble flat space for $k\gg H$. A more exact condition is obtained by setting $k=\bar H$, where $\bar H$ is the value of the Hubble parameter below which the symmetry is restored, as discussed in the next section. This only significantly relaxes the constraint \eqref{eq:alphaconstraint} if $m_\phi\gg\alpha m$, in which case the constraint becomes $\alpha^2\leq\sqrt{18c_{123}\lambda}m_\phi/m$.}
\begin{equation}
\alpha^2\leq54c_{123}\lambda,\label{eq:alphaconstraint}
\end{equation}
where we have assumed $m_\phi\ll\bar\rho$, as we will throughout this paper. This constraint places a mild upper bound on the coupling, $\alpha\lesssim\sqrt\lambda$. We expect $\lambda\lesssim\mathcal{O}(1)$, otherwise the theory is strongly coupled. Only the combination $\alpha m$ is relevant for baryogenesis, and this constraint then implies that we cannot simultaneously take $m<\Mp$ to satisfy the constraints above whilst having $\alpha m$ parametrically larger.

\section{Cosmology}
\label{sec:cosmo}

\begin{figure*}[ht]
{\includegraphics[width=0.45\textwidth]{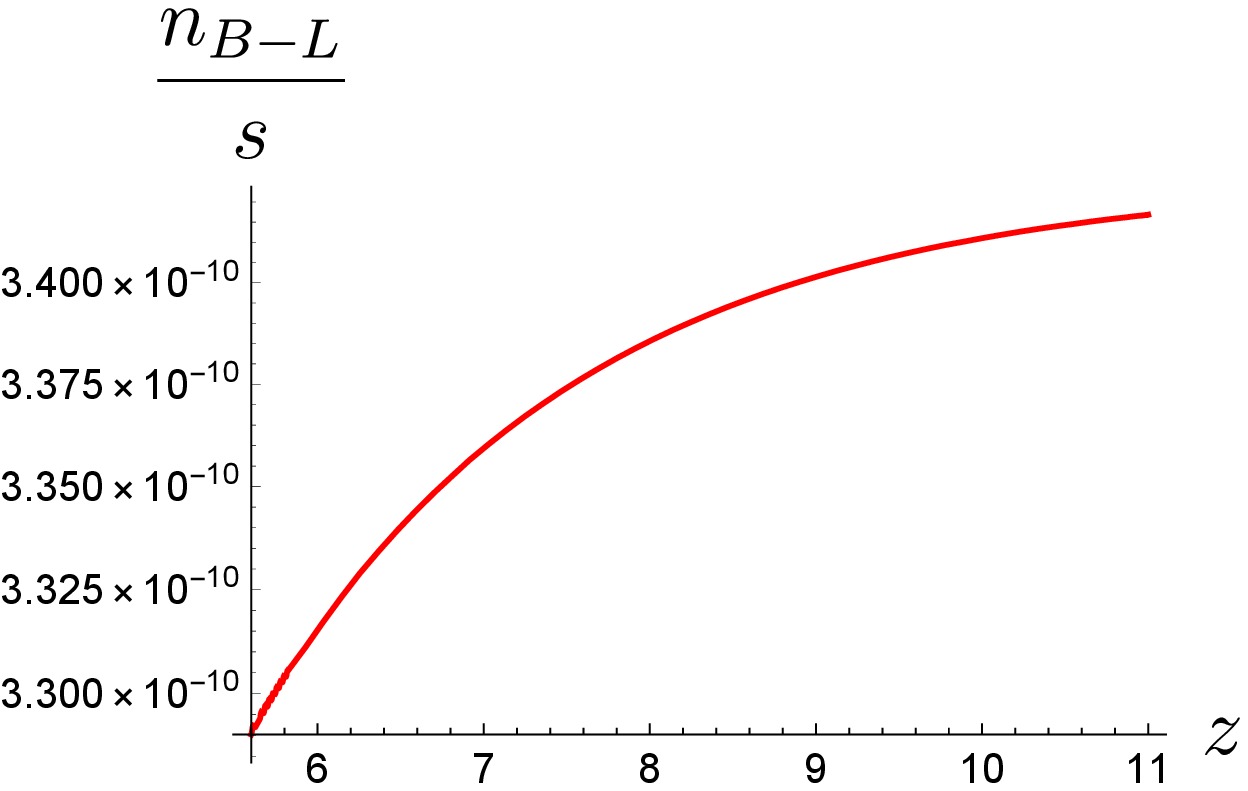}}
{\includegraphics[width=0.45\textwidth]{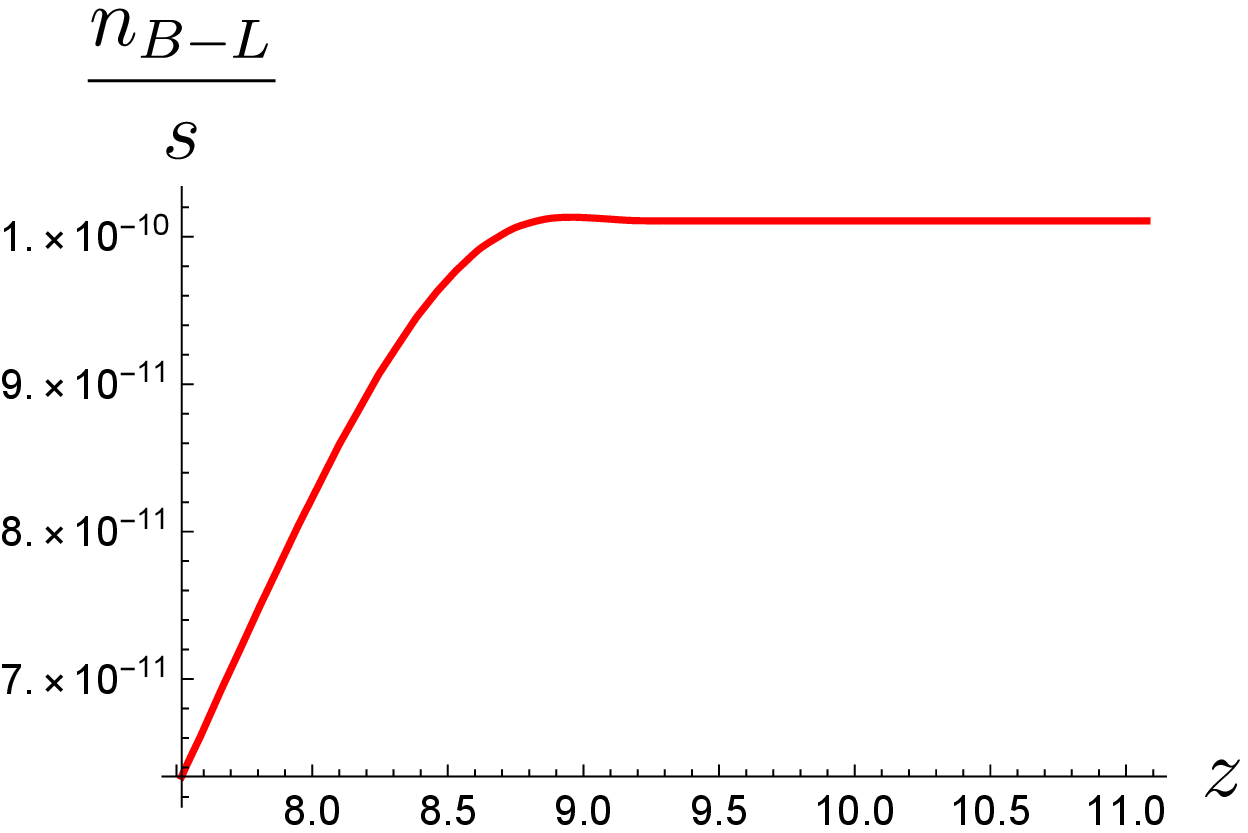}}
\caption{ $\nbl/s$ as a function of $z=\ln(\alpha m t)$. The left panel corresponds to a model with $m_\phi/\alpha m=1/2$, $\alpha m/\Mp=4\times10^{-7}$, and $T_R=10^{11}$ GeV.  The right panel has $m_\phi/\alpha m=7\times10^{-3}$, $\alpha m/\Mp=2\times10^{-6}$, and $T_R=10^{10}$ GeV.}
\label{fig:nbl}
\end{figure*}

In this section we will assume a homogeneous and isotopic cosmological background and derive a simple estimate \eqref{eq:nbls} for the baryon-to-photon ratio generated by our model. In order to verify that the approximations we make are valid, we also solve the equations of motion numerically, with the results plotted in \cref{fig:field,fig:nbl}.

We can see from the potential \eqref{eq:Veff-cosmo} that the $\mathrm{U}(1)$ symmetry is broken at early times and restored at late times, when
\begin{equation}\label{eq:hbar}
H<\bar{H}\equiv\frac{m_\phi^2}{\alpha m}.
\end{equation}
When $H>\bar{H}$ there is a time-dependent symmetry-breaking minimum at
\begin{equation}\label{eq:phimin}
\left|\phi_\mathrm{min}\right|^2=\frac{\alpha m H-m_\phi^2}{\lambda}\approx\frac{\alpha m H}{\lambda},
\end{equation}
where we have assumed that the small symmetry-violating terms ($\varepsilon\phi^4 + \varepsilon^\dagger{\phi^\dagger}^4$) are negligible, or, equivalently, have chosen the coefficients $\varepsilon$ so that this is the case at early times.

The field tracks this minimum nearly adiabatically until $H\approx\bar{H}$, at which point the symmetry is restored and the field begins to oscillate around the symmetry-restored minimum at $\phi=0$. When this occurs, the $\mathrm{U}(1)_{B-L}$-violating terms play an important role. 
We would like these to become important around the time that the symmetry is restored. Expanding $\varepsilon=\bar{\varepsilon}e^{i\psi}$, the correction to the potential is
\begin{equation}
\Delta V_{\rm eff}=\frac12\bar\varepsilon|\phi|^4\cos\left(4\theta+\psi\right),
\end{equation}
where $\phi=|\phi|e^{i\theta}$. The $B-L$ charge density is
\begin{equation}
n_{B-L}=J^0=i(B-L)(\phi^\dagger\overset{\leftrightarrow}{\partial^0}\phi)=2(B-L)|\phi|^2\dot{\theta},
\end{equation}
so we see that the motion of the angular field is responsible for generating a net $B-L$. This means that $\theta$ should not sit at its minimum in the early Universe, and, indeed, one expects it to be frozen at some initial value due to Hubble damping. We would like it to begin rolling around the time of symmetry restoration in order to generate a net $B-L$ before the field settles into the new minimum at $\phi=0$, which will be the case if the canonically-normalized field's mass\footnote{Recall that $\mathcal{L}_\theta/\sqrt{-g}\supset |\phi|^2(\partial\theta)^2$, necessitating the factor of $|\phi|^{-2}$ in the canonical normalization.} $m_\theta^2=\Delta V_{{\rm eff}\,\theta\theta}/|\phi|^2$ is of order $\bar{H}^2$. If $m_\theta<\bar{H}$ the angular field will not roll after symmetry restoration and no $B-L$ will be generated. Similarly, if $m_\theta>\bar{H}$ the field starts rolling long before symmetry restoration, and the value of $B-L$ is set by tuning the initial conditions. Setting $m_\theta\sim\bar{H}$ implies that
\begin{equation}\label{eq:ebar}
\bar{\varepsilon}\sim \lambda\left(\frac{m_\phi}{\alpha m}\right)^2,
\end{equation}
where we have used $|\phi|=|\phi_{\rm min}|$ at the time of symmetry restoration.

Using the angular field's equation of motion,
\begin{equation}
|\phi|^2\left(\ddot{\theta}+3H\dot{\theta}\right)+2|\phi||\dot\phi|\dot{\theta}=\bar{\varepsilon}|\phi|^4\sin(4\theta+\psi),
\end{equation}
one has
\begin{equation}\label{eq:neqn}
\dot{n}_{B-L}+3H\nbl=2(B-L)\bar{\varepsilon}|\phi|^4\sin(4\theta+\psi).
\end{equation}
Making the the approximation $\dot{n}_{B-L}\approx H\nbl$ \cite{Asaka:2000nb,vonHarling:2012yn}, which we will verify numerically later, one finds
\begin{equation}
\nbl\sim\bar{\varepsilon}\frac{|\phi_{\rm min}|^4}{\bar{H}},
\end{equation}
where we have omitted factors of order unity. Using \cref{eq:hbar,eq:phimin,eq:ebar} we can estimate the $B-L$ conserved charge density as
\begin{equation}\label{eq:nbl}
\nbl\sim \frac{m_\phi^4}{\lambda\alpha m}.
\end{equation}

We do not directly observe $\nbl$, but rather the baryon-to-photon ratio $n_\mathrm{b}=\nbl/s$, where $s$ is the entropy density. This introduces some model dependence; for concreteness, and to minimize the number of free parameters, we will focus on a minimal model in which the Universe reheats instantaneously after inflation, and the $\mathrm{U}(1)_{B-L}$ symmetry is restored shortly thereafter. The assumption of instantaneous reheating yields
\begin{equation}
s=\frac{4\rho_I}{3T_R}\sim{\bar{H}^2\Mp^2}{T_R},
\end{equation}
where $T_R$ is the reheat temperature. Combining this with \cref{eq:nbl} we find
\begin{equation}\label{eq:nbls}
n_\mathrm{b} = \frac{\nbl}{s}\sim \frac{10^{-10}}{\lambda}\left(\frac{T_R}{10^8\textrm{ GeV}}\right)\left(\frac{\alpha m}{\Mp}\right).
\end{equation}
We see that this new mechanism can produce the observed baryon-to-photon ratio, $n_\mathrm{b}\sim10^{-10}$, with sensible choices for the reheat temperature and model parameters. Note that the parameters $\alpha$ and $m$ only appear in the combination $\alpha m$, while it is $m^2$ (in combination with the $c_i$) which is constrained by experimental tests of Lorentz violation, as discussed in \cref{sec:constraints}. We expect $\lambda\lesssim\mathcal{O}(1)$, otherwise the theory is strongly coupled, and as discussed above, we should have $\alpha^2\lesssim\mathcal{O}(10)\lambda$ to ensure gradient stability around flat space.

In order to verify the approximations we have made above, we have numerically integrated the scalar field equations assuming a radiation-dominated Universe. In \cref{fig:field} we plot the motion of the complex scalar for two different models. One can see the behavior we predicted qualitatively above: the field tracks its time-dependent minimum at early times before the angular field begins to roll when the symmetry is restored, giving rise to a spiral trajectory. In \cref{fig:nbl} we plot the baryon-to-photon ratio $\nbl/s$ for the same models. One can see that our numerical results agree well with our prediction \eqref{eq:nbls}.

\section{Conclusions}
\label{sec:conc}

In this paper we have studied baryogenesis in Lorentz-violating theories of gravity, which, at low energies, are naturally described by a constrained vector so that there is a preferred frame. Baryogenesis requires a field charged under $\mathrm{U}(1)_{B-L}$, the simplest choice being a complex scalar. We have demonstrated here that the lowest-order interaction between the scalar and vector can give rise to a tachyonic mass term for the scalar proportional to the Hubble parameter so that the $\mathrm{U}(1)_{B-L}$ symmetry is broken at early times. Inverse phase transitions such as these can generate a net $B-L$ through the coherent motion of the scalar when the symmetry is restored at late times, and we have shown here that Lorentz-violating theories can successfully generate the observed baryon-to-photon ratio using this phenomenon. Furthermore, this can be achieved for parameter choices that are not ruled out by current constraints. Our theory differs from the quintessential paradigm---the Affleck-Dine mechanism---in that the tachyonic mass is proportional to $H$ rather than $H^2$, which gives rise to new features and a qualitatively different cosmology, which we have calculated in detail. Lorentz-violating gravity theories continue to be important in the study of dark energy and quantum gravity. Here, we have shown that they may also shed light on the origin of the mater-antimatter asymmetry.   

\acknowledgements

We are grateful to Mark Trodden for enlightening discussions. JS and ARS are supported by funds provided to the Center for Particle Cosmology by the University of Pennsylvania. This paper is dedicated to the memory of Little Pete's diner, 1978--2017. \textit{Benedictio caro cum caseo sit.}

\bibliography{ref}
\bibliographystyle{apsrev4-1}

\end{document}